\documentclass[reprint,aps,prl]{revtex4-1}
\usepackage{graphicx}
\usepackage{bm}
\usepackage{color}
\usepackage{subfigure}
\usepackage{ulem}
\usepackage{amsmath}
\usepackage{braket}
\usepackage{hyperref}

\begin{document}
\title{Imaging the Holon String by Quantum Interference} 
\author{Tin-Lun Ho}
\affiliation{Department of Physics, The Ohio State University, Columbus, OH 43210, USA} 
\date{\today}
\begin{abstract} 
It has been a  long sought goal of Quantum Simulation to find answers to long standing questions in condensed matter physics.  A famous example  is the ground and the excitations of 2D Hubbard model with strong repulsion 
below half filling.  The system is a doped antiferromagnet. It is of great interests because of its possible relation to high Tc superconductor. Theoretically, the fermion excitations of this model are believed  to split up into holons and spinions, and a moving holon is believed to  leave behind it a string of ``wrong" spins that mismatch with the antiferromagnet background.    Here, we show that the properties of the ground state wavefunction and the holon excitation of the 2D Hubbard model can be revealed in unprecedented detail using the technique of quantum interference in atomic physics. This is achieved by using quantum interference to measure the Marshall sign of the doped antiferromanget. The region of wrong Marshall sign directly reflects the spatial extent of fluctuating string attached to the holon. 
\end{abstract}
\maketitle

Fermi Hubbard model is among the most important models in condensed matter physics. It is  exceedingly simple -- a set of fermions in a lattice with local interaction. Yet it is  notoriously difficult to solve. Its two dimensional  (2D) version  with repulsive interaction $U>0$ is particularly famous because of  its relevance to high $T_{c}$ superconductivity\cite{Anderson}.  Recently, this 2D model has been realized in optical lattices using ultra-cold fermions, and its antiferromagnetic correlations has been observed\cite{Greiner, Bloch, Martin, Tilman}. Due to the flexibility in controlling density, interaction, and lattice parameters in cold atom experiments, it is  hoped that many long standing questions about this model such as the nature of  ground state and the mechanisms for charge and spin transport can be answered.  Recently, there have been  experimental studies of these transports for the 2D Hubbard model\cite{Bakr, Martin-spin-diffusion}. However, the microscopic origins of the observed properties still have to be understood. 

In the strong repulsion limit, the ground state of the Hubbard model at half filling is an antiferromagnet (AF). Each lattice site is occupied by a fermion. The  problem of central interest is the nature of the ground state when a density of holes is introduced in the AF.  The current view is that fermion excitations of this system are made up  holons and spinons interacting with gauge fields between them\cite{Lee}.  A holon carries charge but no spin, whereas a spinon carries spin but no charge. This is very different from the excitations of a Fermi liquid, which carry both charge and spin. Mathematically, a holon and a spinor are defined through the so called slave boson method\cite{slaveboson}.  Often, a holon is represented pictorially as a hole in a classical AF, $|AF\rangle_{c}$,  which has   $\uparrow$ and $\downarrow$ spins occupying the sublattice A and B  respectively.  With this classical approximation of the AF,  a moving hole will leave behind it a string of wrong spins that cannot be healed by nearest neighbor spin exchange.  (See Figure 1(a) and 1(d)). This string will cost magnetic energy and is expected to hinder the motion of the hole.  However, this description does not include quantum fluctuations, which are known to be important as they lead to a large reduction of the fermion magnetic moment. With quantum fluctuations, the  concept of a string is less well defined, as  different fluctuation configurations will lead to different string patterns. See Figure 1. 

\begin{figure}[htbp]
\centering
\includegraphics[width=3.5in, height=2.0in]{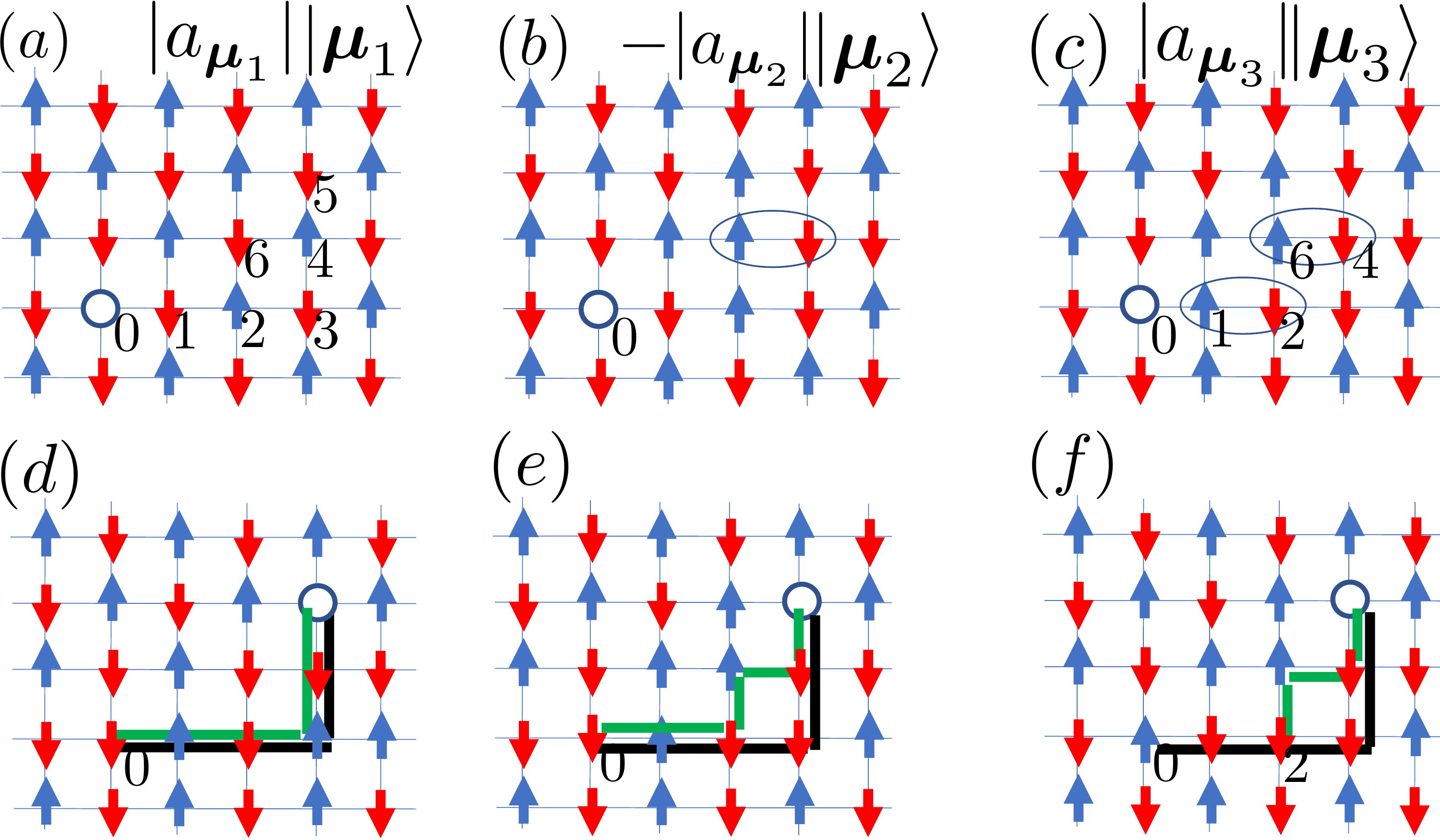}
\caption{The ``classical procedure" for identifying holon strings and the effects of quantum fluctuations:  The AF ground state $|F\rangle_{\bf 0}$ with a hole fixed at site-0 consists the classical AF spin configuration (a), as well as quantum fluctuations  (with less weight) such as (b) and (c) obtained from exchanging pairs of neighboring opposite spins in (a). 
Figures (d)-(f) are the spin configurations when the hole hops from 0 to 5 along the path 12345  (black line) starting from  (a) to (c).  Figure (d) is the usual description of the holon string, generated by a hole moving in a classical AF background. Within the classical procedure, the sequence of mismatched spins (i.e. those on the green line in (d)) follows the trajectory of the hole.  For (e) and (f), if the hole slides back along the green path, we end up with a classical AF state with a hole at 0 and 2 respectively.  When taking a  snap short of the spin density after the hole has traveled from 0 to 5, one   might pick up  spin configurations such as (e) and (f). However, the classical procedure will arrive at  two different string patterns ( i.e.  the green lines)  even both are generated by the same motion of the hole.  Moreover, neither string follows the actual trajectory.  
Returning to (a)-(c), the AF state with an immobile hole at 0,   the classical procedure will imply that (c) arises from having a hole moved from site-4 to site-0 in a classical AF through the path 6210, whereas the hole has never moved at all. 
To illustrate another effect of quantum fluctuations, we return to (a)-(c),  the AF state with an immobile hole at 0.  The classical procedure will conclude (c) is the string pattern generated by a hole moving from site-4 to site-0 along the path 6210 in a classical AF, even the hole has never moved at all. 
All these show that quantum fluctuations will cause ambiguities in the classical procedure. 
The spin states of  (a), (b), and (c) are denoted as $|\bf{\mu}_{1}\rangle $, $|\bf{\mu}_{2}\rangle $  and $|\bf{\mu}_{3}\rangle $, their amplitudes in the in the ground state $|F\rangle_{0}$ are  $+|a_{\bm{\mu}_1}|$, $-|a_{\bm{\mu}_2}|$, and $+|a_{\bm{\mu}_3}|$, where the signs are specified by the  Marshall sign rule.}
\label{Marshall-1}
\end{figure}

Recently, Markus Greiner's group has tried  to identify the holon strings  in a doped AF by comparing the experimentally observed  spin configuration to that of a classical AF\cite{H-string}. This scheme, which we refer to as the 
``classical" procedure", has not included quantum fluctuations and the rotational invariance of the AF state. 
In this paper, we point out a method to identify the holon string that is free from these problems. 
Our method is to use quantum interference technique to measure the Marshall sign of the AF. We shall see that for the AF state with an immobile hole in (Figure 1(a)-1(c)), the Marshall sign for all neighboring sites is $(-1)$.  On the other hand, if the hole is allowed to move. The Marshall sign in the immediate vicinity of the trajectory  will change to +1, due to a path dependent phase caused by the AF background.  Since the distribution of Marshall sign has included all quantum fluctuations of the AF and respects its rotational invariance, it constitutes an operational definition of the holon string.

{\em The Heisenberg model and the Marshall sign:} 
It is well known that the  Hubbard model is $H= - t\sum_{<i,j>, \sigma}c^{\dagger}_{\sigma}(i) c^{}_{\sigma}(j) + U \sum_{i} n_{\uparrow}(i) n_{\downarrow}(i)$ in the large repulsion limit  ($U\gg t>0$) reduces to the 
$tJ$ model $H_{tJ}$ which  operates on the  space of no double  occupancy.  The model is 
$H_{tJ}= {\cal T} + H_{J}$, 
\begin{equation}
H_{J} =  J \sum_{<i,j>} {\bf S}_{i}\cdot {\bf S}_{j},   \,\,\,\,\,\, J= t^2/U>0
\end{equation}
\begin{equation}
{\cal T}=  -t  \sum_{<i,j>,  \sigma} \overline{c}^{\dagger}_{\sigma}(i) \overline{c}^{}_{\sigma}(j), \,\,\,\,\,\,  \overline{c}_{\sigma}(i) = c_{\sigma}(i) (1- n_{-\sigma}(i)),   
\end{equation}
where $n_{\sigma}(i) = c^{\dagger}_{\sigma}(i)
c^{}_{\sigma}(i)$,  ${\bf S}_{i} = c^{\dagger}_{\mu}(i) \bm{\sigma}_{\mu\nu} c^{}_{\nu}(i)/2$ , 
$H_{J}$ is the AF Heisenberg hamiltonian 
 between the spins at nearest neighbors $( i, j )$,   and ${\cal T}$ is the hopping of a fermion to its neighbor. 
At half filling, each site contains exactly one fermion, hence  ${\cal T}=0$ and  $tJ$ Hamiltonian reduces to $H_{J}$. The  quantum states are of the form 
$|\Psi\rangle = \sum_{\bm{\mu}} \Psi(\bm{\mu})|\bm{\mu}\rangle$, 
where $\bm{\mu}\equiv  (\mu_1, \mu_2, \mu_3, ...\mu_{N})$ stands for 
the spin configuration with spin $\mu_{i}$ at site ${\bf R}_{i}$ (denoted simply as $i$); $\mu =1 (-1)$ for  $\uparrow$ ($\downarrow$); $|\bm{\mu}\rangle = c^{\dagger}_{\mu_1}(1)c^{\dagger}_{\mu_2}(2) .. c^{\dagger}_{\mu_N}(N)|0\rangle$; $N$ is total number of lattice sites, and $\Psi(\bm{\mu}) \equiv   \Psi \left(  1,\mu_{1}\, ;  2,\mu_{2}\, ;  ... ; N,\mu_{N}\, \right)$. 
If the system has a hole {\em fixed} at site-$i$,  the spin basis will then be denoted as 
$ |\bm{\nu}; i\rangle$, where $\bm{\nu}$ represents the spin configuration on all  $(N-1)$ occupied sites. The quantum state will then be written as $|\Psi; i\rangle = \sum_{\bm{\nu}} \Psi(\bm{\nu}; i) 
|\bm{\nu}; i\rangle$. In both cases,  $H_{tJ}$ reduces to $H_{J}$. 

In 1955, Marshall  showed  that  the {\em ground} state  $|F\rangle$ of the  Heisenberg model on a bipartite lattice (with sublattice A and B) will change sign if a pair of opposite spins at nearest neighbor sites $(i,j)$ are interchanged\cite{Marshall}, i.e.
\begin{equation}
 \Psi( i \uparrow; j \downarrow; ..) = - \Psi( i \downarrow; j \uparrow; ..) 
 \label{sign} \end{equation}
 where $\Psi( i \uparrow; j \downarrow; ..) \equiv  \Psi( .. ; i \uparrow; ... ;  j \downarrow; ...)$, and $(...)$ means all other  spins. Marshall  also showed  that the sign rule  Eq.(\ref{sign}) means the ground state $|F\rangle$ is of  the form\cite{Marshall},  
\begin{equation}
F(\bm{\mu}) = (-1)^{N^{\downarrow}_{A}(\bm{\mu})} \overline{F}(\bm{\mu}), \,\,\,\,\,\,\  \overline{F}(\bm{\mu})\geq 0, 
\label{zeta} \end{equation}
where $N^{\downarrow}_{A}(\bm{\mu})$ is the total number of down spin in the sublattice $A$ in the configuration $\bm{\mu}$. This is because interchanging two opposite spins in two nearest neighbor sites with change $N^{\downarrow}_{A}$ by 1, as the two sites must belong to different sublattices.  This motivates  one to define a spin basis that obeys the Marshall sign for any number of immobile holes\cite{Weng, Weng2}, 
\begin{equation}
 \overline{ |\bm{\mu}\rangle} = (-1)^{N^{\downarrow}_{A}} | \bm{\mu}\rangle, \,\,\,\, 
 \overline{  |\bm{\mu}; i\rangle} = (-1)^{N^{\downarrow}_{A}} | \bm{\mu}; i\rangle . 
  \end{equation}
Expanding a state in this basis $|\Psi \rangle = \sum_{\bm{\mu}} \overline{\Psi}(\bm{\mu}) \overline{ | \bm{\mu}\rangle}$,  $|\Psi; i\rangle = \sum_{\bm{\mu}} \overline{\Psi}(\bm{\mu}; i) \overline{ 
| \bm{\mu}; i\rangle}$, 
the condition for satisfying the Marshall sign rule Eq.(\ref{sign}) is that the coefficients 
$\overline{\Psi}(\bm{\mu})$ (or $\overline{\Psi}(\bm{\mu}, i)$) are  real positive numbers for all spin configurations (apart from an overall phase which we ignore). 

Before proceeding, we give a proof of the Marshall sign rule for Heisenberg models on bipartite lattices with any fixed number of holes. It is a 2D consequence of the Perron-Frobenius theorem, which says that a real square matrix with non-negative entries has a unique largest eigenvalue, and that the corresponding eigenvector has all positive components. For the Heisenberg model $- \sum_{\langle i,j\rangle} ({\bf S}_{i}\cdot {\bf S}_{j} - 1/4)$ with $Q$ immobile holes at ($i_1, i_2, ..., i_Q$), it is easy to verify that the matrix element in the basis  $\overline{ |\bm{\mu}; i_1, .. i_Q\rangle} =  (-1)^{ N_{A}^{\downarrow}} |\bm{\mu}; i_1, .. i_Q\rangle$ are all positive. The Marshall sign rule for the ground state of $H_{J}$ then follows from the  Perron-Frobenius theorem for the highest energy state of $-H_{J}$. 

{\em Measurement of the Marshall sign:}
To detect the  relation Eq.(\ref{sign}),  one needs to  interfere two of its wavefunctions differing only by  an exchange of opposite spins at nearest neighbors  $(i,j)$. Such interference is contained in 
 the ``exchange overlap"  $\rho_{ij}= \langle c^{\dagger}_{\downarrow}(i)  c^{\dagger}_{\uparrow}(j)  c^{}_{\downarrow}(j)  c^{}_{\uparrow}(i) \rangle_{\Psi}$, 
\begin{eqnarray}
\rho_{ij}^{} 
= \sum_{(..)} \Psi( i \downarrow;  j \uparrow; ..)^{\ast} \Psi(i \uparrow;  j\downarrow; ..)
\hspace{1.0in}  \nonumber \\
= - \sum_{(..)} \overline{\Psi}( i\downarrow;  j \uparrow; ..)^{\ast} \overline{\Psi}( i \uparrow;  j \downarrow; ..)  \equiv |\rho_{ij}^{} | e^{i\theta},   \hspace{0.2in} 
\label{rhoij} \end{eqnarray}
where $\theta$ and $e^{i\theta}$ will be referred to as Marshall angle and Marshall sign respectively. 
If $|\Psi\rangle$ is the ground state $|F\rangle$, Eq.(\ref{sign}) implies $\theta =\pi$ and a Marshall sign -1. 

To create the function $\rho_{ij}$, we can first perform  separate spin rotations on the fermions at nearest neighbor sites $i$ and $j$, 
($U = U_{i} U_{j})$, and then measure the  correlations, say, of the up-spins, i.e. 
\begin{equation}
\langle n_{\uparrow}(i)n_{\uparrow}(j)\rangle_{\Psi'}= \langle \Psi| U^{\dagger}n_{\uparrow}(i) n_{\uparrow}(j)U|\Psi\rangle
\label{mix} \end{equation}
where $|\Psi'\rangle = U|\Psi\rangle$. The spin rotations will  mix $\uparrow$ and $\downarrow$ spin at each site $i$ and $j$.  The correlation in Eq.(\ref{mix}) will then pick up the interference term of opposite spins in $i$ and $j$. Explicitly, 
under a spin rotation $U$, a spin  transforms as $U^{\dagger}c^{\dagger}_{\uparrow}U = u c^{\dagger}_{\uparrow} + v c^{\dagger}_{\downarrow}$, $U^{\dagger}c^{\dagger}_{\downarrow}U= -v^{\ast} c^{\dagger}_{\uparrow} + u^{\ast} c^{\dagger}_{\downarrow}$, where $|u|^2 + |v|^2 =1$. 
 If $|\Psi\rangle$ is a spin eigenstate (i.e. fixed $S_z$),  and if we take  $(u_{i}, v_{i}) = (1,1)/\sqrt{2}$, $(u_{j}, v_{j}) = (1,e^{i\beta})/\sqrt{2}$, then we have 
 \begin{equation}
W(\beta) \equiv \langle n_{\uparrow}(i)n_{\uparrow}(j)\rangle_{\Psi'}^{} - \frac{1}{4} = 
 \frac{1}{2}|\rho_{ij}| {\rm cos}(\theta - \beta). 
\label{W} \end{equation}
Repeating  the   measurement $\langle n_{\uparrow}(i)n_{\uparrow}(j)\rangle_{\Psi'}$ 
for different $\beta$, one can obtain the interference term  $W(\beta)$ and back out $\theta$ from its maximum.  

{\em Motion of a hole in an antiferromagnetic background:} 
In the following, we shall study the motion of a single hole in an AF background. 
Similar study has been performed numerically on a 4-leg cylinder focusing on long time behavior\cite{F-4-legs}.  The strings of holons have  has also been studied in terms a parton model \cite{F-parton-neel, F-mag-polaron}, which had been used to discuss the string pattern in ref.\cite{H-string} deduced from the classical procedure.  
Here, we shall apply  our formulation of Marshall sign basis to obtain analytic and exact results for the holon strings for time intervals below 1/t. 
We shall assume that we can reach temperatures low enough so that the spins are essentially  in the ground state of the Heisenberg hamiltonian.  We start with the AF ground state
$|F, {\bf 0}\rangle$ with a hole fixed at ${\bf R}={\bf 0}$ (site-0). This state can be created by a  strong blue detuned  laser focused at ${\bf 0}$. $|F, {\bf 0}\rangle$ has the expansion 
$|F; {\bf 0}\rangle = \sum_{\bm{\nu}} \overline{|\bm{\nu}; {\bf 0}\rangle} \, \,\overline{F_{\bf 0}}(\bm{\nu})$,  with $\overline{F_{\bf 0}}(\bm{\nu})  >0$, 
where  $\bm{\nu}$ denotes the configuration of the rest $L-1$ spins, and
$\sum_{\bm{\nu}} |\overline{F_{\bf 0}}(\bm{\nu})|^2=1$. If the focused laser is suddenly removed at time $\tau=0$, the hole will hop according to the hamiltonian $H_{tJ}$. If the hole is found at site-${\bf R}$ after time $\tau$, then the system is in the (un-normalized) state  $|\Phi(\tau) \rangle =   \sum_{ \bm{\nu}}  \overline{|\bm{\nu}; {\bf R} \rangle} \,\, \overline{  \Phi}_{\bf R}(\bm{\nu}; \tau)$,
where 
\begin{equation}  
\overline{\Phi}_{\bf R}(\bm{\nu}; \tau)  = \sum_{\bm{\mu}_{0}} \overline{  \langle \bm{\nu}, {\bf R}|} \,  e^{-i \tau H_{tJ}} \, \overline{  |\bm{\mu}_{\bf 0}, {\bf 0} \rangle  }\,\,\,  \overline{F_{\bf 0}}(\bm{\mu}_{\bf 0}). \label{Phij}  \end{equation} 
Although $\overline{F_{\bf 0}}$ carries the Marshall sign, $\overline{\Phi}_{\bf R}$ need not be due to the propagator. 
Evaluating the propagator in Eq.(\ref{Phij}) is a formidable task. However, the calculation can be  simplified when $t\gg J$, which is satisfied in current experiments. In this limit, over the time interval  $\tau < 1/t \ll 1/J$, the spins hardly evolve and can be treated perturbatively. To the lowest order of $J/t$, 
the time revolution can be replaced by $e^{-i{\cal T}\tau}$. 

 \begin{figure}[htbp]
\centering
\includegraphics[width=3.0in, height=1.6in]{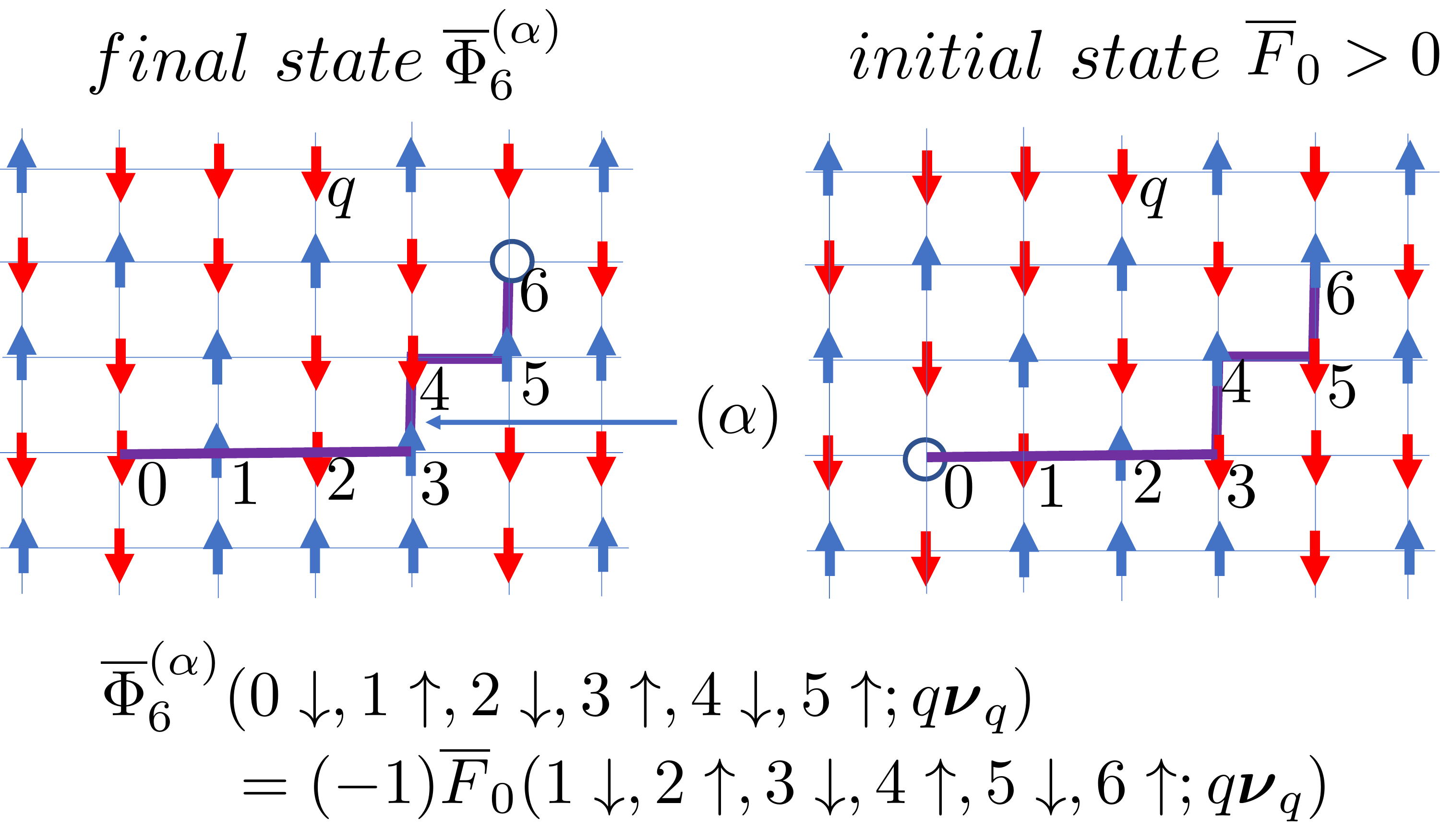}
\caption{
 Starting with an AF ground state with a hole at ${\bf R}_{0}$, the  initial wavefunction $\overline{F}_{0}$ is positive for all spin configurations due to the Marshall sign rule. After the hole hops from 0 to 6 (or ${\bf R}_0$ to $R_{6}$ through the path
$(\alpha) = (0,1,2,..,6)$, 
The wavefunction  $\overline{\Phi}_{6}^{(\alpha)}$ is related to 
$\overline{F}_{0}$ as $\overline{\Phi}_{6}^{(\alpha)}(0\nu_{0}, 1\nu_1, ..., 5\nu_5;  q \bm{\nu_{q}})= \overline{F}_{0}( 1\nu_0, 2\nu_1, ..,  6\nu_5;  q \bm{\nu_{q}}) \zeta(\bm{\nu}^{(\alpha)})$,  
where $q$ labels the sites off the path $\alpha$. For the spin patterns in the figure, we have $\bm{\nu}^{(\alpha)} = (\nu_{0}, \nu_1, ..., \nu_5)=  (\downarrow, \uparrow, \downarrow, \uparrow, \downarrow, \uparrow)$; and $\zeta (\bm{\nu}^{(\alpha)})= \nu_{0}\nu_{1} \nu_{2} \nu_{3}  \nu_{4}\nu_{5}=-1$.
} 
\label{fig:n=3cabbage}
\end{figure}

The  minimum number of hops from {\bf 0} to ${\bf R}$ is  $n = |R_x| + |R_y|$. 
There are $(|R_x| + |R_y|)!/(|R_x| !  |R_y| !)$ paths with $n$ hops, each of which (labeled as $\alpha$) is a sequence of sites $(\alpha)= (0,1,2,..n) =  ({\bf R}_{0}, {\bf R}_{1}, ..., {\bf R}_{n})$, 
where ${\bf R}_{0}={\bf 0}$ and ${\bf R}_{n}={\bf R}$. The wavefunction in Eq.(\ref{Phij}) is then a sum over different paths,  $\overline{\Phi}_{\bf R}= \sum_{\alpha} \overline{\Phi}^{(\alpha)}_{\bf R}$. 
Ignoring an overall constant $(it\tau)^{n}/n!$, we have 
\begin{eqnarray}
\overline{\Phi}^{(\alpha)}_{ {\bf R}}(\bm{\nu}; \tau) 
= \sum_{[\bm{\mu}]} \prod_{j=1}^{n} \overline{ \langle \bm{\mu}_{j}; {\bf R}_{j} | } \, (-{\cal T}/t) \, \overline{ | \bm{\mu}_{j-1}, {\bf R}_{j-1}\rangle}^{(\alpha)} 
\nonumber \\
 \times \overline{F}_{0}(\bm{\mu}_{0})  + O((t\tau)^2)  \hspace{0.7in}
\label{Phial} \end{eqnarray}
where $\bm{\mu}_{n} = \bm{\nu}$, and the sum is over the spin configurations $(\bm{\mu}_0, \bm{\mu}_1, \bm{\mu}_{n-1} )$ for different location of the hole before it reaches ${\bf R}$. 
The matrix element $\overline{ \langle \bm{\nu} ;  \ell |} \, {\cal T} \,  \overline{| \bm{\nu}'; \ell'\rangle}$,     is non-vanishing  only when the sites $\ell$ and $\ell'$ are nearest neighbors, and with spin configurations differing only by a  transfer of a single spin associated with the hopping hole\cite{Weng}. Explicitly, we have 
\begin{equation}
\overline{ \langle \bm{\nu} ;  \ell |} \, (-{\cal T}/t) \,  \overline{| \bm{\nu}'; \ell'\rangle}  =   \nu_{\ell'} \,  \delta(\nu_{\ell'}, \nu'_{\ell}) \prod_{ q \neq \ell, \ell'} \delta(\nu_{q}, \nu'_{q}). 
\label{hop} \end{equation}
The spin dependent phase $\nu_{\ell'}$ in Eq.(\ref{hop}) is the essence of hole hoping in the AF background. Without it, the  amplitude would be identical to that of a free fermion.

 To evaluate Eq.(\ref{Phial}) for a given path $(\alpha)$, we divide the spins $\bm{\nu}$ into the set  on the path (denoted as $\bm{\nu}^{(\alpha)}$) and off the path (denoted as $\tilde{\bm{\nu}}^{(\alpha)}$);  $\bm{\nu} = \bm{\nu}^{(\alpha)} \oplus \tilde{ \bm{\nu}}^{(\alpha)}$. 
  The spins $\tilde{ \bm{\nu}}^{(\alpha)}$ outside the path are not affected by the hopping of the hole, whereas those on the path simply shift down by one step as the hole hops from ${\bf 0}$ to 
  ${\bf R}$.  See Figure 2.   In other words, if ($\nu_{0}, \nu_{1}, ..\nu_{n-1}$) are the spins 
  at (${\bf R}_{0}, {\bf R}_{1}, ... {\bf R}_{n-1})$ when the hole arrives at ${\bf R}_{n}={\bf R}$ through path
  $(\alpha)$, then it is originated from the initial state (where the hole is at ${\bf R}_{0}= {\bf 0}$) with the same set of spin 
  ($\nu_{0}, \nu_{1}, ..\nu_{n-1}$) located at  (${\bf R}_{1}, {\bf R}_{2}, ... {\bf R}_{n})$.  At the same time, Eq.(\ref{hop}) implies a path dependent phase due to the AF  spin background 
   \begin{equation}
\zeta(\bm{\nu}^{(\alpha)}) = \nu_0 \nu_1 .. \nu_{n-1} = (-1)^{N_{\downarrow}(\bm{\nu}^{(\alpha)})}, 
\label{sp} \end{equation}
where $N_{\downarrow}(\bm{\nu}^{(\alpha)})$ is the number of down spins on the path $(\alpha)$. 
Due to this phase factor, ref.\cite{Weng} refers the path $(\alpha)$ as a ``phase-string".   Here, we show that to the lowest order of $J/t$,  the  amplitude  $\overline{\Phi}^{(\alpha)}_{\bf R}$ on this path is given by the ground state amplitude as 
\begin{equation}
\overline{\Phi}_{{\bf R}}^{(\alpha)}(\bm{\nu}; \tau) =  \zeta(\bm{\nu}^{(\alpha)})\, 
 \overline{ F_{\bf 0}} (\bm{\nu}^{(\alpha)};  \tilde{\bm{\nu}}^{(\alpha)}).
\label{ftau} \end{equation}

\begin{figure}[htbp]
\centering
\includegraphics[width = 2.5in, height=2.0in]{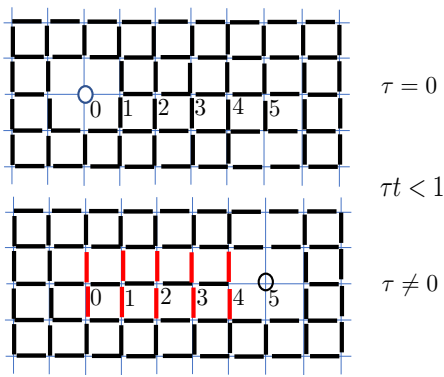}
\caption{The distribution of Marshall signs after the hole has traveled from 0 to 5 at time $\tau$ such that $\tau t<1$:  The initial state (upper figure) is an AF ground state with a hole fixed at site-0.  The Marshall sign $e^{i\theta}$ for any pair of nearest neighbor sites is  $-1$ (represented by a thick black line).  For 
$\tau t<1$, the dominant path going from 0 to 5 is the straight line that connects them. All other  paths have amplitudes  at least $(\tau t)^{2}$ smaller. For this dominant path, the distribution of Marshall sign is shown in the lower figure. The Marshall sign at the immediate vicinity of the path changes from -1 to +1 (represented by a thick red line). 
}
\label{fig:n=3cabbage}
\end{figure}

To find  the Marshall sign of  the wavefunction Eq.(\ref{Phij}),  let us consider the simple case where site-${\bf R}$ is $n$ steps away from ${\bf 0}$ along $\hat{x}$. In this case, there is only one  path connecting ${\bf 0}$ and ${\bf R}$ to the lowest order of $t\tau$;  i.e. the straight line  $(\alpha)=(0,1,2,..n)$. See Figure 3.   Eq.(\ref{Phij}) then contains a single term.   The exchange overlap  (Eq.(\ref{rhoij})) for neighboring sites $(i,j)$ is 
\begin{eqnarray}
\rho_{ij} = - \sum_{(..)} \overline{ F_{0}}(i \downarrow; j \uparrow; ...)
\overline{ F_{0}}(i \uparrow; j \downarrow; ...)   \nonumber \\
\times 
(-1)^{N_{\downarrow}(\bm{\mu}_{ij}^{(\alpha)}) + N_{\downarrow}(\bm{\mu}_{ji}^{(\alpha)}) }
\hspace{0.3in} 
\label{line} \end{eqnarray}
where $\bm{\mu}_{ij}$ denotes the spin configurations where the spins at  site-$i$ and $j$ are fixed at $\uparrow$ and $\downarrow$, i.e. $\bm{\mu}_{ij}= (i \uparrow; j \downarrow; ...)$;  $(...)$ denotes the spins at all other sites;  and $\bm{\mu}_{ij}^{(\alpha)}$ denotes those spins on the straight line path $(\alpha)$. 
If both  $i$ and $j$ are  on the path or off the path $(\alpha)$,   then $N_{\downarrow}(\bm{\mu}_{ij}^{(\alpha)})= N_{\downarrow}(\bm{\mu}_{ji}^{(\alpha)})$, and hence $\rho_{ij}^{\overline{\Phi}} = \rho_{ij}^{\overline{F}}<0$. 
 This means $\theta = \pi$, and a Marshall sign -1 for all  nearest neighbor pairs $(i,j)$, exactly the same as  the AF ground state. 
If $i$ is on the path and $j$ is off the path, then
$N_{\downarrow}(\bm{\mu}_{ij}^{(\alpha)})$ and $N_{\downarrow}(\bm{\mu}_{ji}^{(\alpha)})$ differ by 1,
we then have  $\rho_{ij}^{\overline{\Phi}} = -\rho_{ij}^{\overline{F}} > 0$. We then have  $\theta = 0$ and 
a Marshall sign +1 for the nearest neighbor sites  in the immediate vicinity of the hole trajectory as shown in Figure 3. 
A measurement of the distribution of the Marshall sign then maps out the trajectory of the holon, and the region of Marshall sign violation defines the string attached to the hole. 

If ${\bf 0}$ and ${\bf R}$ are not on the same symmetry axis, there are more than one path that connect them.  The exchange overlap Eq.(\ref{rhoij}) of the state in Eq.(\ref{Phij}) 
is of the form $- \sum_{\alpha, \beta} \overline{\Phi_{\bf R}}^{\alpha}\overline{\Phi_{\bf R}}^{\beta}$. The off diagonal terms with $(\alpha\neq \beta)$ are  weaker than the diagonal ones (with $\alpha=\beta$)  as the phase fluctuations of different paths do not cancel.  Including only the diagonal terms, we have 
$\rho_{ij} = \sum_{\alpha} \rho_{ij}^{(\alpha)} /\sum_{\alpha} 1$. 
Since each path will lead to a violation of Marshall sign in it vicinity, and since all the paths  converge at the starting and end site-${\bf 0}$ and site-${\bf R}$, the sign violation will be maximum in the neighborhood of these sites.  These general cases with be discussed elsewhere. 

{\em Experimental scheme for detecting the holon string}:  One immediate question is that after the hole is release, it will go anywhere after time $\tau$.  In order to make use of our analytic results, we need to fix the final position of the hole. This can be done by post-selection of data as follows: 
(A) Starting with an initial ground state with a hole  fixed at ${\bf R}={\bf 0}$, one release the hole at time $\tau=0$ by suddenly removing the potential that creates the hole. 
(B) After time $\tau < 1/t$, one performs the spin rotations at  neighboring sites $i$ and $j$  with angle $\beta$ as discussed in the text and then images of the spin density immediately.   
(The spin rotations are to prepare for the construction of the function $W(\beta)$). This process is repeated for a large number of times,  $M\gg 1$. Note that the probability for the hole to travel $n$-steps is $(t\tau)^{n}$.  
 (C)  One repeats step (B) for different values of $\beta$. (D) Among the images for each $\beta$, one selects $Q$ ($M> Q\gg 1$) images where the hole ends up at site ${\bf R}$ along the $x$-axis. After averaging over these $Q$ images, we obtain the interference function $W(\beta)$ in Eq.(\ref{W}) for any nearest neighbor sites $i$ and $j$, from which one can back out the Marshall sign.  The region where the Marshall sign is violated then maps out the holon string.  Of course, to determine the Marshall sign of all nearest neighbor pairs will require very large number of measurements. However, this number can cut down significantly if one focus only in the neighborhood of the straight line connecting ${\bf 0}$ and ${\bf R}$ as shown in Figure 3. 

{\em Further Remarks}: 
Fermi Hubbard model is a major focus of Quantum Simulation.
Here, we present a method to reveal a fundamental property (the Marshall sign) of its AF phase, which can be applied to track the motion of a hole  and to identify the string attached to it.  
This method can be generalized to  multi-holes, with spin and doublon fluctuations treated within perturbation theory. Our results show that atomic physics experiments are  powerful new ways to reveal  the fundamental properties of strongly correlated systems, and will help unravel the mysteries of doped antiferromagnets. 

 Acknowledgments: The work is supported by the MURI Grant FP054294-D and the NASA Grant on Fundamental Physics 1541824. This work was completed  during a visit at the IAS of HKUST in January 2012,  I thank Professor Gyu-Boong Jo and Director Andy Cohen for hospitality and arrangements. 


\end{document}